# Simple Fuzzy Score for Russian Public Companies Risk of Default

By
Sergey Ivliev[1]

April 12, 2010

## 1. Introduction

Current economy crisis of 2008-2009 has resulted in severe credit crunch and significant NPL rise in Russian financial system. Bad debt rate has almost tripled to 6% of total loans to corporate sector, while provisions reached 9%. Corporate bonds market cannot be called safe heaven either having almost 25% of issuers defaulted on their debt. Probably the only good side of it is that the researchers and the banks can now have a better sample to identify and test predictive power of their credit risk models. The number of default cases including public defaults on bonds market is now significantly higher than in previous years.

This paper describes a score model constructed to give a reasonable and simple way of measuring the risk of default. The model is aimed to discriminate the «good» and the «bad» companies in Russian corporate sector based on their financial statements data (Russian Accounting Standards based). The validation shows performance at about 72% in-sample Gini accuracy ratio and pretty good match to external ratings. The model provides a very simple rule to estimate implied credit rating for Russian companies which many of them don't have.

## 2. Data collection

*Data Sample.* The data sample consists of 126 Russian public companies- issuers of Ruble bonds which represents about 36% of total number of corporate bonds issuers. 25 companies have defaulted on their debt in 2008-2009 which represents around 30% of default cases. 29% companies in the sample have credit ratings assigned compared to 34% in the parent population. No SPV companies were included in the sample.

The source of financial statements data: [Prime-Tass Database](#).

*Definition of "Bad" case and Cure events.* The case (reporting date) is treated as "bad" when the company first defaults on its debt after the report is made available public. According to Russian legislation 30 days lag was used for quarterly reports (1Q,2Q,3Q), 90 days lag – for annual report, e.g. if default event happened on 17/02/2009, the report issued not later than the 3Q 2008 should be treated as a "bad" case. Only real defaults are analyzed. Cure events (technical defaults) are treated as "good".

---

[1] JSC Prognoz, Perm State University. Email: ivliev@prognoz.ru

The source of defaults data: TRUST Interactive: Defaults Review

*Observation period.* Observation period for the model is 1Q 2008 – 3Q 2009. Total number of cases (company*quarters) is 588 (4.6 data points per company on average).

*Missing values.* Missing values in the source financial statements data were case-wise excluded from the sample.

*Data processing.* Financial statements data were processed to calculate financial ratios commonly used in corporate creditworthiness analysis:

| Size | Balance sheet structure | Profitability | Liquidity |
|---|---|---|---|
| LN (Assets) | Working Capital / Assets | EBIT / Sales | Cash / ST Debt |
| LN (Sales) | Retained Earnings / Assets | EBIT / Assets | Cash and equivalents/ ST Debt |
| Sales/Assets | Equity / Total Liabilities | EBIT / Interest | Current Assets/ ST Debt |

Altman's Z"-Score (EM Score) was also calculated.

*Predictive power.* For each variable discriminating power was analyzed based on in-sample Gini accuracy ratio. Best predictors were identified with the power above 50% as follows: Equity to Total Liabilities ratio, EBIT / Interest coverage ratio, Scale of company's Sales and Retained Earnings to Asset ratio.

| Variable | In-sample Gini AR |
|---|---|
| Equity / Total Liabilities | 57.8%. |
| EBIT / Interest | 55.8% |
| LN(Sales) | 54.3% |
| Retained Earnings / Assets | 52.3% |

While the other variables shows much less and even negative predictive power:

| Variable | In-sample Gini AR |
|---|---|
| EBIT / Assets | 39.6% |
| Sales / Assets | 23.3% |
| EBIT / Sales | 10.6% |
| Altman's Z"-Score (EM Score) | -7.7% |

ROC-curves of best predictors are shown on fig.1.

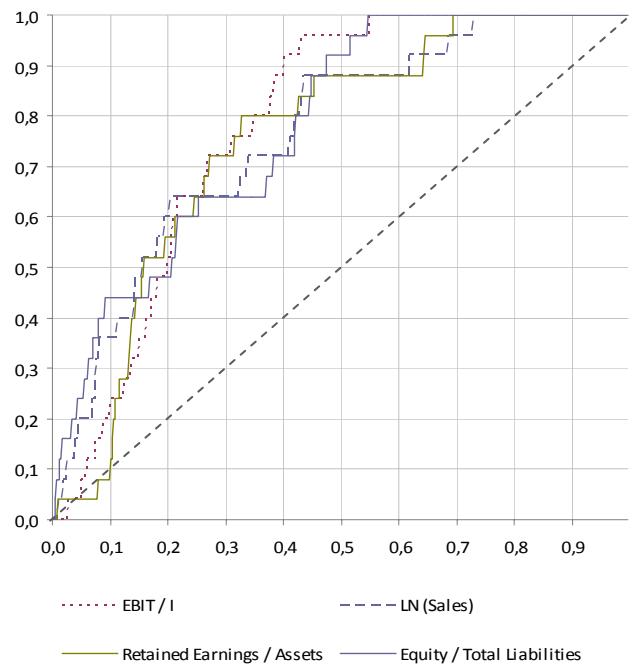

Fig.1. ROC-curves of best predictors.

*Correlations.* Correlation matrix was estimated for best predictors to avoid multicollinearity risk.

|                      | LN (Sales) | Ret Earnings / Assets | Equity / Liabilities |
|----------------------|------------|-----------------------|----------------------|
| EBIT / I             | 0,1160     | 0,3015                | 0,3726               |
| LN (Sales)           |            | 0,2104                | 0,3507               |
| Ret Earnings / Assets |           |                       | 0,133                |

Scatter plot diagrams for each pair of the variables are shown below (see fig.2) to make sure that no collinearity effects are present.

Having the first principal component yielding less than 50% of the variance we can state that there is no need to use any dimension reducing techniques such as factor analysis or principal components method.

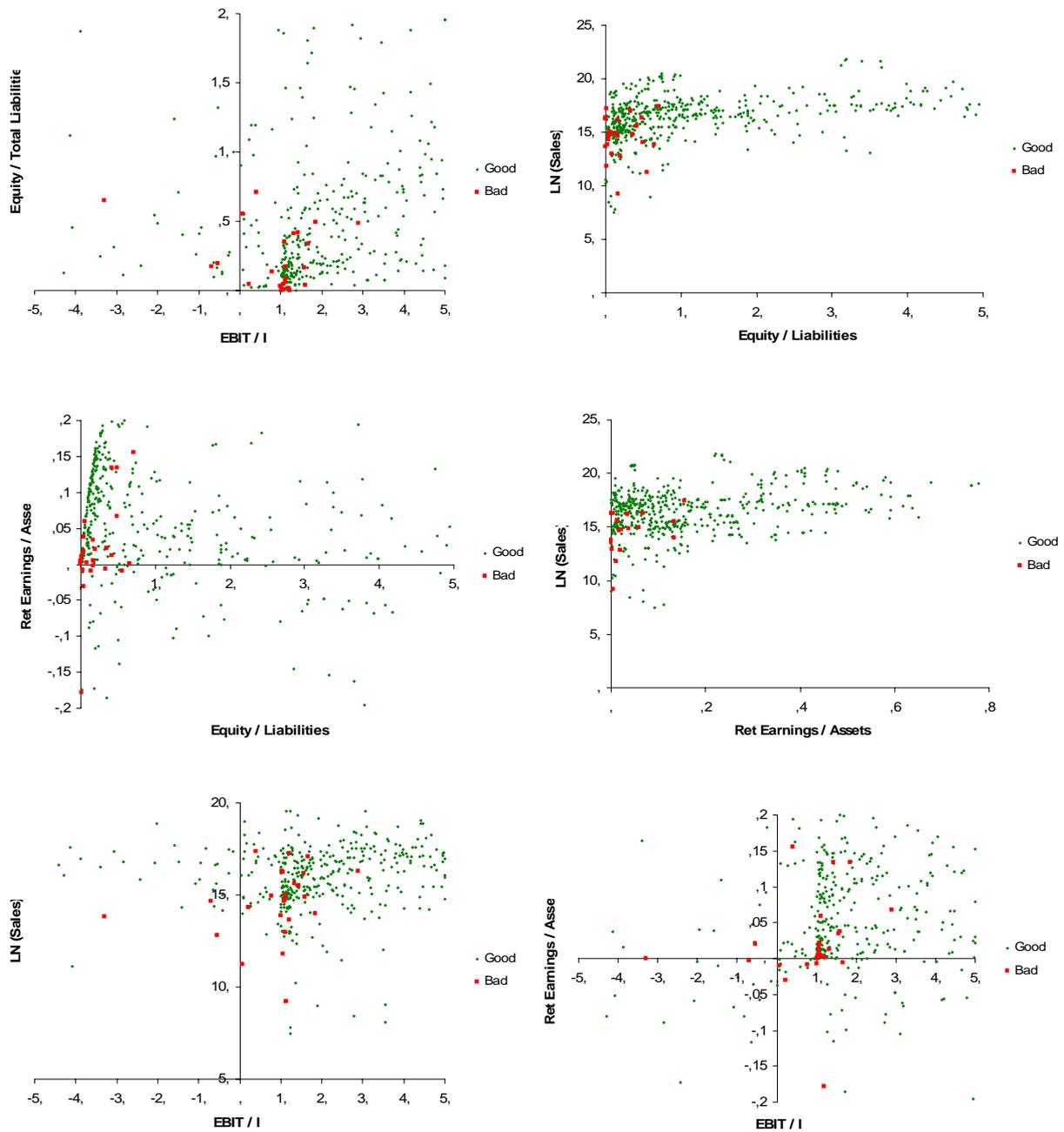

Fig.2. Scatter plot diagrams of predictors

## 3. Prediction model design

*Predictors empirical distribution analysis.* Empirical CDFs show that predictive variables are tending to be asymmetric and heavy tailed (see fig.3). This fact sets a certain limitations on estimation techniques that require normality of variables such as MDA.

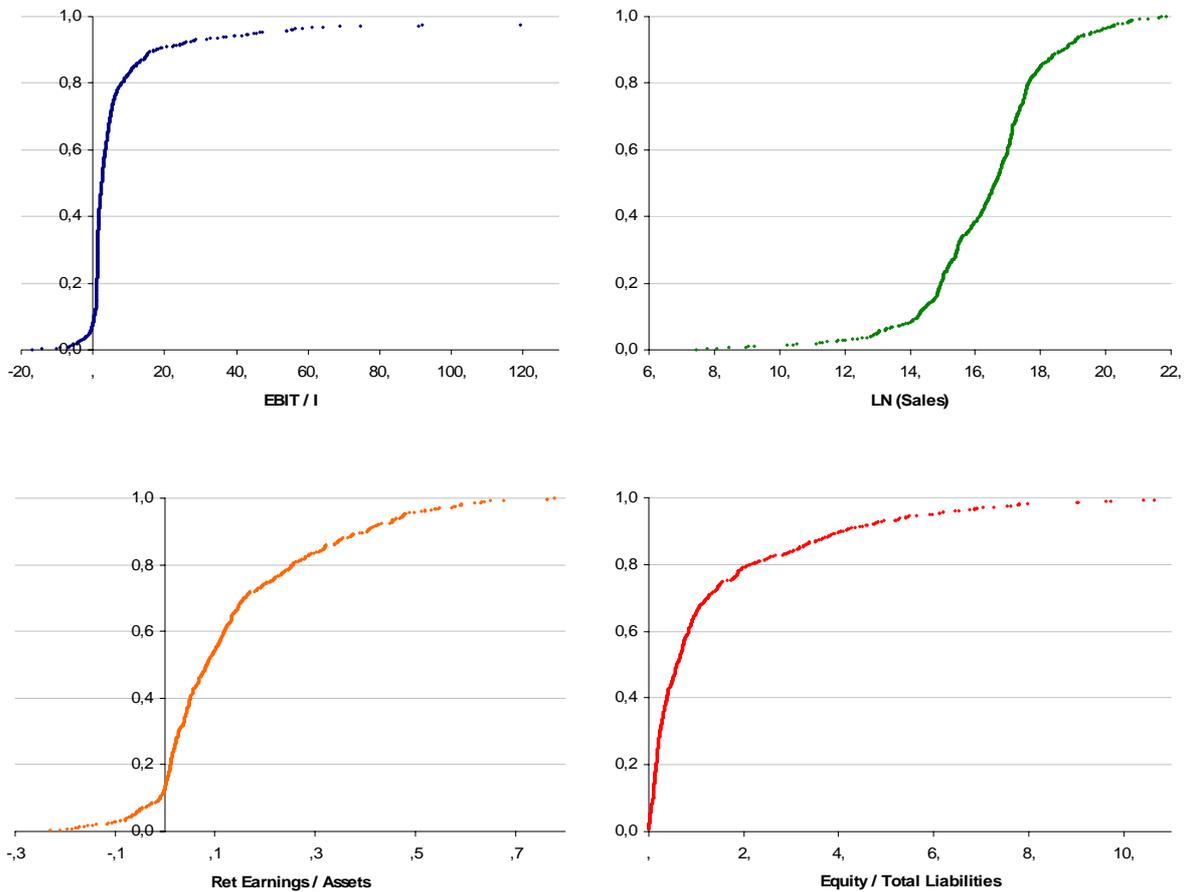

Fig 3. Empirical distributions of predictors

*Methodology choice.* The general requirements for methodology are reasonable to be the following:

1. The method should be simple enough to be easily explained to non-professionals. The extra complexity of the method only makes sense when extra predictive power is gained;
2. The method should be robust enough to handle non Gaussian distributions;
3. The method should take advantage of good explanatory power of variables.

The following models were chosen to be validated:

1. Simple cut-off model (*S-Score*);
2. Simple fuzzy score model (*FS-Score*);
3. Logistic model (*Logit*);
4. Logistic model based on fuzzy variables (*Logit F*).

The simplest cut-off model (*S-Score*) can be formalized as:

$$S = \sum_{i=1}^{n} 1_{\{x_i > c_i\}}$$

For each predictor the best cut-offs $c_i$ were estimated to provide the smallest total misclassification error (type I + type II error).

| Variable | Best cut-off | Type I error | Type II error | Total error |
|---|---|---|---|---|
| EBIT / Interest | 2 | 44% | 4% | 48% |
| LN(Sales) | 16 | 37% | 28% | 65% |
| Retained Earnings / Assets | 0.04 | 33% | 20% | 53% |
| Equity / Total Liabilities | 0.5 | 44% | 12% | 56% |
| **S-Score** | **1** | **27%** | **8%** | **35%** |

The S-Score has provided in-sample Gini AR equal to 71.8%.

In order to make the model score more continuous the fuzzy sets approach with the linear membership functions.

$$\gamma_i(X_i, a_i, b_i) = \begin{cases} 0, & X_i < a_i \\ \frac{X_i - a_i}{b_i - a_i}, & a_i \leq X_i < b_i \\ 1, & X_i \geq b_i \end{cases}$$

The $a_i$ cut-offs were set the same as in *S-Score* to discriminate the "bad" while the $b_i$ were set to provide the highest membership grades for excellent corporations.

| Variable | $a_i$ cut-off | $b_i$ cut-off |
|---|---|---|
| EBIT / Interest | 2 | 7 |
| LN(Sales) | 16 | 18 |
| Retained Earnings / Assets | 0.04 | 0.2 |
| Equity / Total Liabilities | 0.5 | 2 |

Conditional CDFs and fuzzy set membership functions are shown in fig.4.

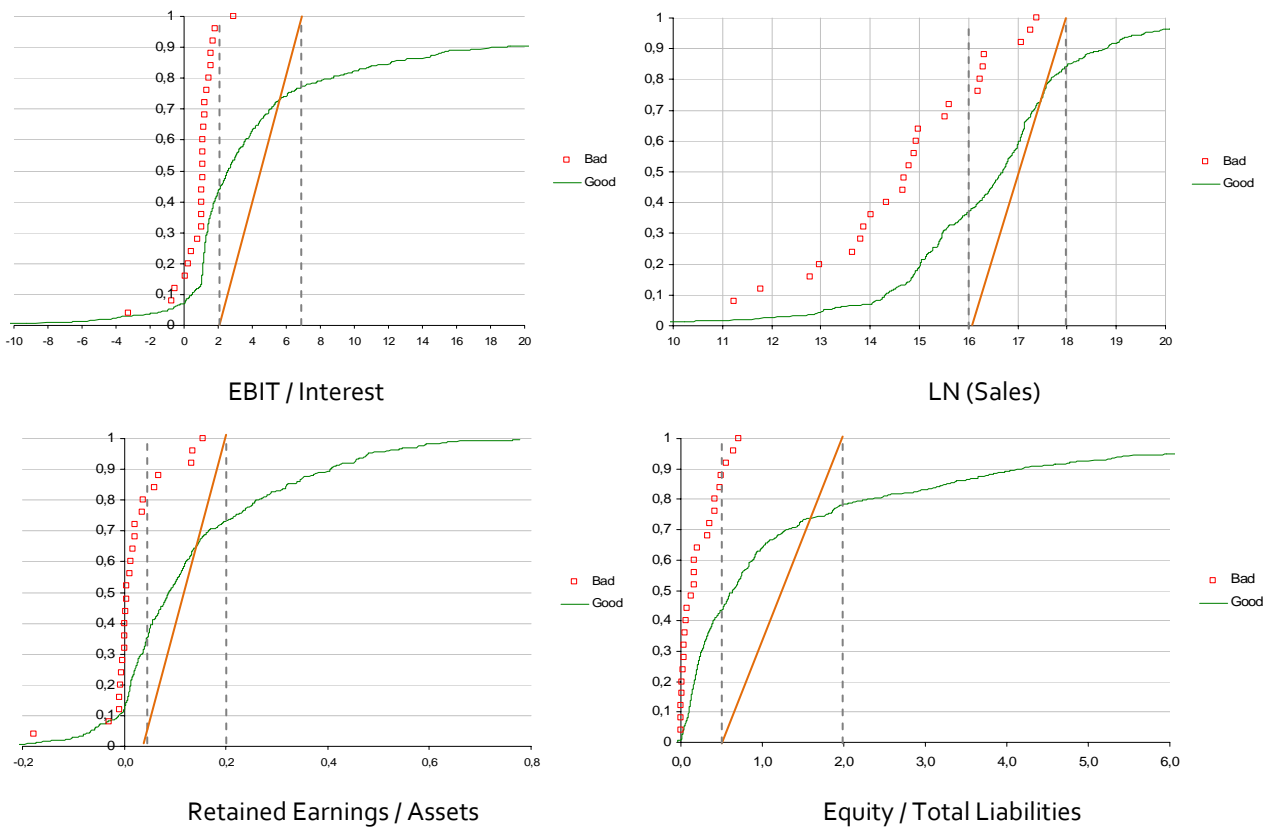

EBIT / Interest

LN (Sales)

Retained Earnings / Assets

Equity / Total Liabilities

A simple sum (*FS-Score*) is used to integrate scores:

$$FS = \sum_{i=1}^{n} \gamma_i(X_i, a_i, b_i)$$

The FS-Score has made better in-sample Gini AR = 72.7%, and provides continuous distribution of the model scores (see fig.).

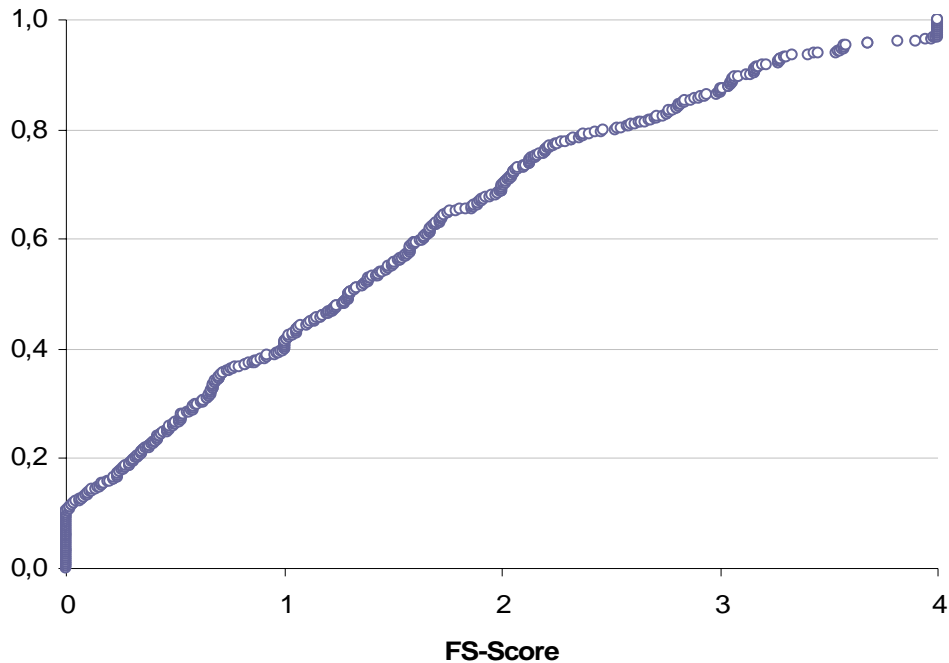

Fig 5. Empirical distributions of predictors

The logistic regression was also tried in order to deliver a better fit. Two logit models were estimated based on initial financial ratios and on fuzzy transformed.

The binary logit on initial financial ratios:

$$y = \frac{e^{b_0 + \sum_{i=1}^{n} b_i X_i}}{1 + e^{b_0 + \sum_{i=1}^{n} b_i X_i}}$$

where y- binary variable (0 for "good", 1 for "bad"), $X_i$ – predicting variable, $b_i$ – regression coefficients.

The estimates of coefficients are shown below:

| Variable | i | $b_i$ |
|---|---|---|
| Const | 0 | 1,9808 |
| EBIT / Interest | 1 | -0,1131 |
| LN(Sales) | 2 | -0,2431 |
| Retained Earnings / Assets | 3 | -3,1491 |
| Equity / Total Liabilities | 4 | -2,0711 |

The in-sample Gini Accuracy Ratio: 70.5%.

Logit model based on fuzzy transformed variables was also estimated:

$$y = \frac{e^{b_0 + \sum_{i=1}^{n} b_i \gamma_i}}{1 + e^{b_0 + \sum_{i=1}^{n} b_i \gamma_i}}$$

where y - binary variable (0 for "good", 1 for "bad"), $\gamma_i$ – fuzzy membership functions value for variable i, $b_i$ – regression coefficients.

The estimates of coefficients are shown below:

| Variable | i | $b_i$ |
|---|---|---|
| Const | 0 | -1,46645 |
| EBIT / Interest | 1 | -6,21185 |
| LN(Sales) | 2 | -1,19298 |
| Retained Earnings / Assets | 3 | -3,1798 |
| Equity / Total Liabilities | 4 | -5,09643 |

The in-sample Gini Accuracy Ratios: 72.9%.

The ROC-curves are shown on fig.6.

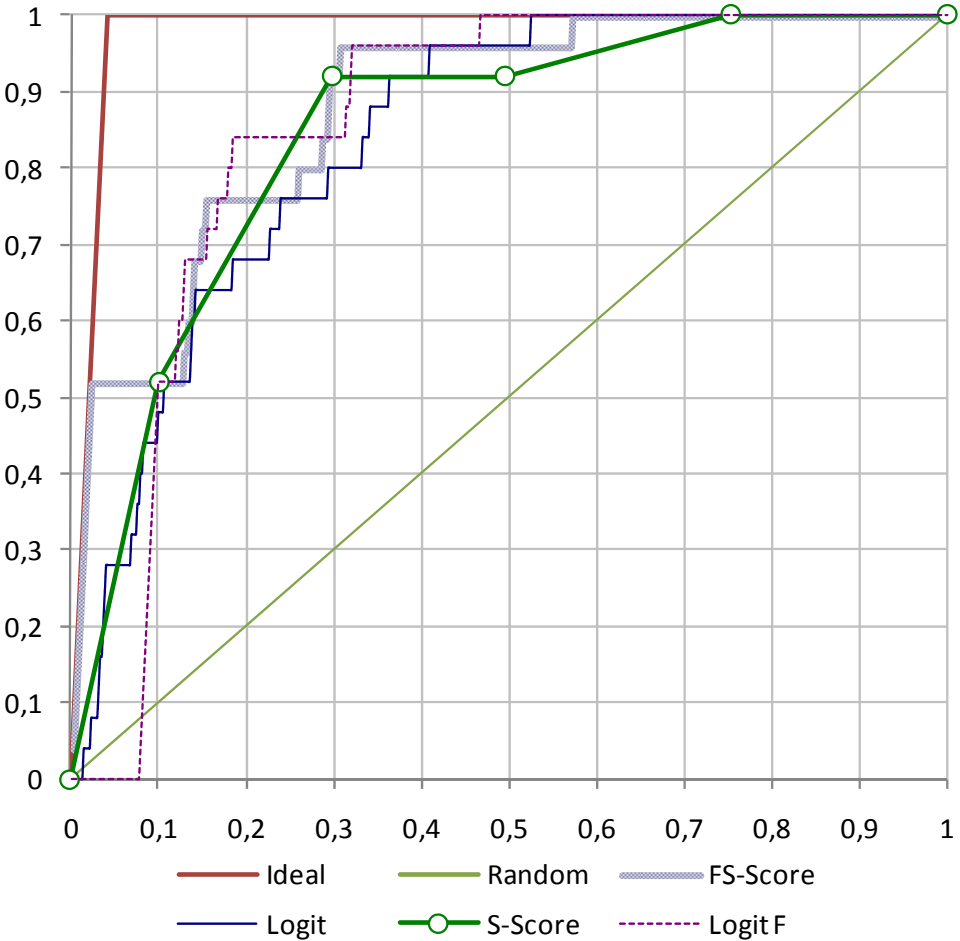

Fig 6. ROC-curves of models

Models look quite similar in terms of their predictive power. *FS-Score* and *Logit F* are a bit more powerful. While *FS-Score* is also more simple.

## 4. External Ratings Calibration

In order to create an internal rating system we calibrated the *FS-score* model to map to external ratings which were assigned by major agencies (S&P, Moody's, Fitch).

The external ratings scale was reduced to 5 major grades: A, BBB, BB, B, CCC/C.

*FS-Score* distribution functions are shown separately for each of the grade (see fig.7).

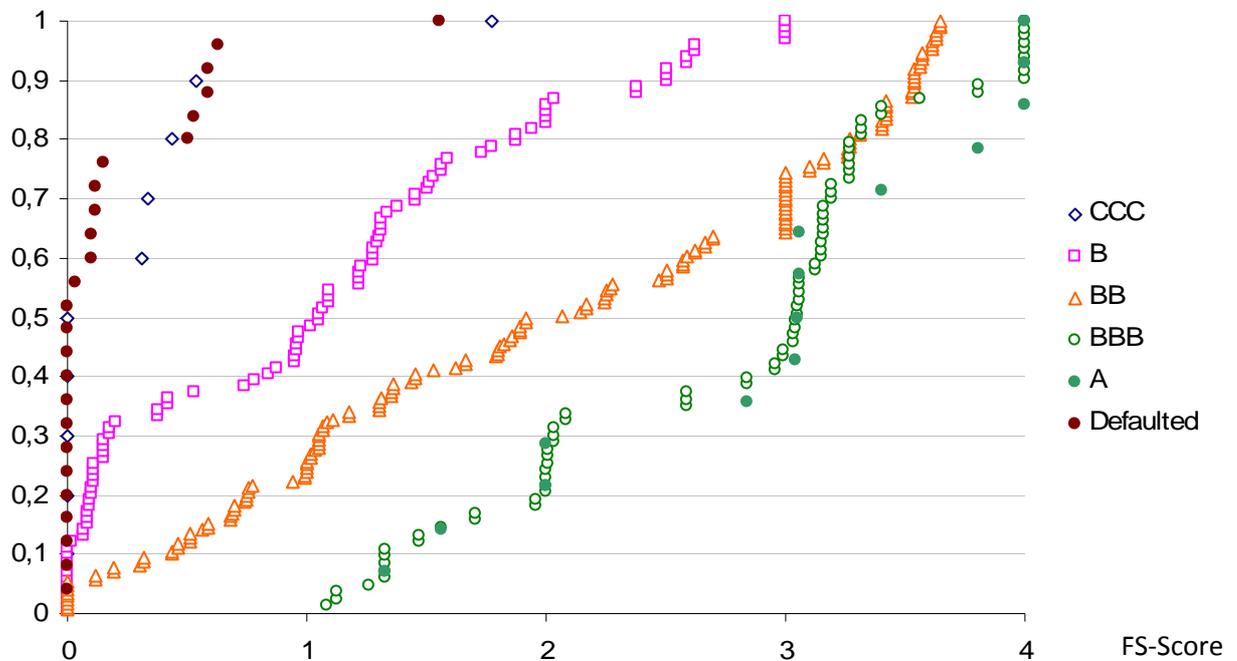

Fig 7. FS-Score distribution among ratings groups

Basic statistics is given below:

| Rating grade | 25% Quintile | Median | 75% Quintile |
|---|---|---|---|
| A | 2 | 3.06 | 3.81 |
| BBB | 2.01 | 3.05 | 3.27 |
| BB | 1 | 2.11 | 3.1 |
| B | 0.11 | 1.05 | 1.56 |
| CCC/C | 0 | 0.15 | 0.44 |
| Defaulted | 0 | 0 | 0.15 |

The medians (except for A's and BBB's) are significantly different and could be interpreted as the centers of the classes. So internal ratings cut-off's might be set just in between the medians.

| Internal rating grade (fs) | Left cut-off | Center | Right cut-off |
|---|---|---|---|
| fsBBB | 2.5 | 3 | 3.5 |
| fsBB | 1.5 | 2 | 2.5 |
| fsB | 0.4 | 1 | 1.5 |
| fsCCC/C | 0.075 | 0.15 | 0.4 |
| fsD | 0 | 0 | 0.075 |

**A simple rule can be applied here: *FS-Score* equals the number of B letters in external rating.** E.g. *FS-Score* close to 2 corresponds to BB.

## 5. Summary

The *FS-Score* model constructed in this paper on one hand gives a good explanation of the defaults of Russian public companies in 2008-2009, while on the other hand being rather simple to be used in wide risk-management practice as a fast risk measure.

The *FS-Score* model is benchmarked to external ratings scale letting the use of external Probabilities of Default (PD) from agencies transition matrices. A very simple rule can be applied to value the credit grade of unrated company: *FS-Score equals number of B's in external rating*.

There are ways of further model validation and improvement:

1. Out-of sample validation based on most recent defaults in Q4 2009 and Q1 2010;
2. Extending the set of predictors based on financial statements and other sources;
3. Analyzing stability of model across economy sectors and in time;
4. Extending the predictive power of the score model with the use of copula function;
5. Using differentiated weights for Type I and Type II errors while estimating the cut-offs of predictors to capture the different effects on losses;
6. Implementation of model for estimation the fair price of bonds.